\documentclass[letterpaper]{jpconf}
\usepackage{graphicx}
\usepackage[symbol]{footmisc}
\renewcommand{\thefootnote}{\fnsymbol{footnote}}
\usepackage{amsmath}

\usepackage{citesort}
\usepackage{amssymb}
\usepackage{fancyhdr}
\begin{document}

\title{Scale-invariance, dynamically induced Planck scale
and inflation in the Palatini formulation}

\author{Ioannis D.~Gialamas$^{1}$, Alexandros Karam$^2$, Thomas D.~Pappas$^3$, Antonio Racioppi$^2$ and Vassilis C.~Spanos$^1$}

\address{$^1$ National and Kapodistrian University of Athens, Department of Physics, Section of Nuclear {\rm \&} Particle Physics,  GR--157 84 Athens, Greece}
\address{$^2$  Laboratory of High Energy and Computational Physics, 
National Institute of Chemical Physics and Biophysics, R{\"a}vala pst.~10, Tallinn, 10143, Estonia}
\address{$^3$  Research Centre for Theoretical Physics and Astrophysics, Institute of Physics, Silesian University in Opava, Bezručovo nám.~13, CZ-746 01 Opava, Czech Republic}

\ead{i.gialamas@phys.uoa.gr, alexandros.karam@kbfi.ee, thomas.pappas@physics.slu.cz, antonio.racioppi@kbfi.ee, vspanos@phys.uoa.gr}

\begin{abstract}
We present\footnote{Talk given by I.D.~Gialamas at HEP 2021, Thessaloniki, Greece} two scale invariant models of inflation in which the addition of quadratic in curvature terms in the usual Einstein-Hilbert action, in the context of Palatini formulation of gravity, manages to reduce the value of the tensor-to-scalar ratio. In both models the Planck scale is dynamically generated via the vacuum expectation value of the scalar fields.
\end{abstract}

\section{Introduction}
In physical cosmology, cosmic inflation~\cite{Guth1981,Starobinsky1982,Linde1982,Albrecht1982} is a theory which describes a
period of exponential expansion of space in the early Universe. The theory
of inflation manages to simultaneously solve basic issues of the Big Bang
cosmology like the horizon and flatness problems.
The simplest theory of inflation assumes the existence of one scalar field $\phi$ which is minimally coupled to gravity and has a canonical kinetic term.

\begin{figure}[h]
\centering
\includegraphics[width=0.8\linewidth]{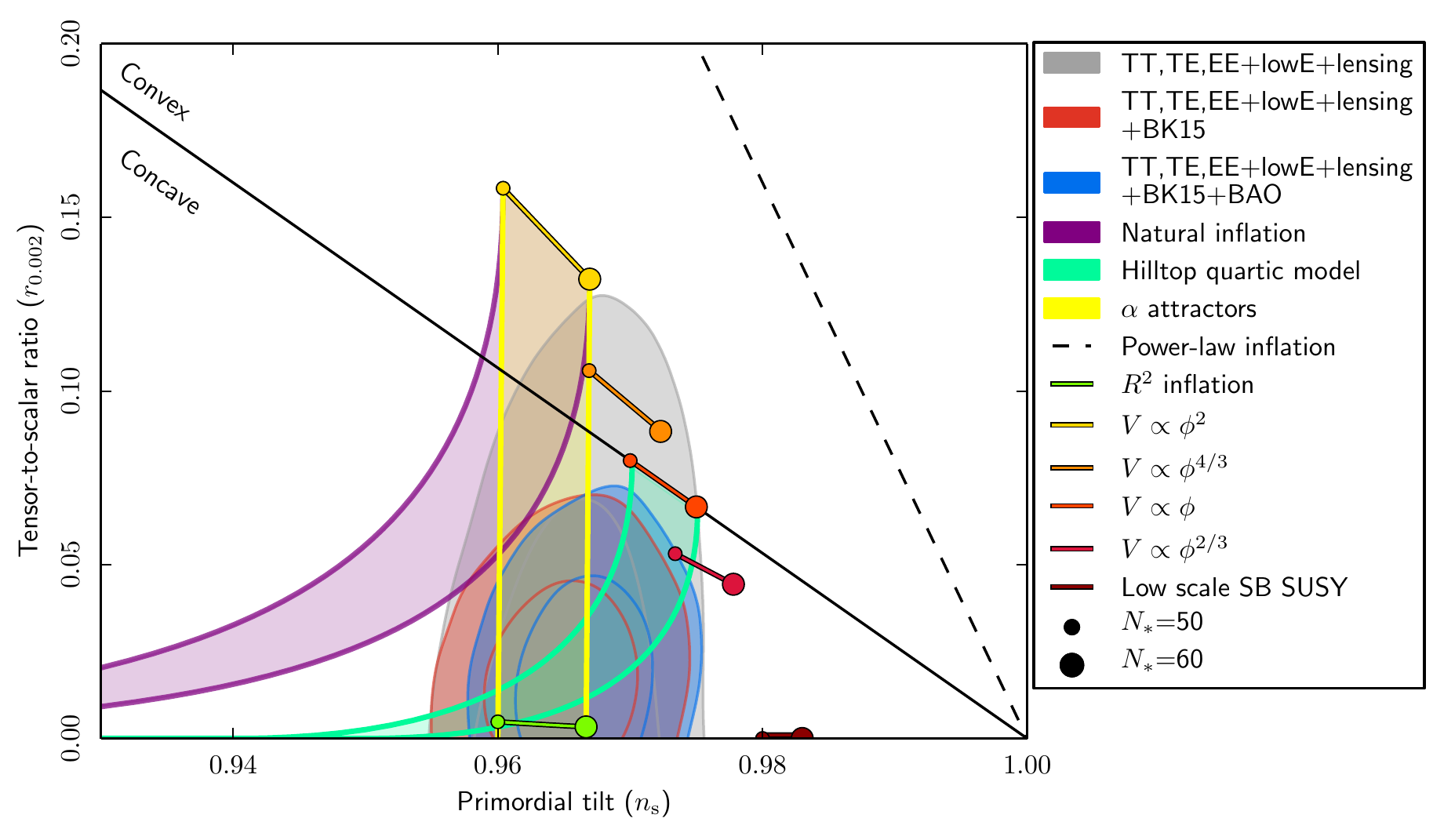}
\caption{Marginalized joint $68 \% $ and $ 95 \% $ CL regions for $n_s$ and $r$ at $k = 0.002\, Mpc^{-1}$  from Planck alone and in combination with BK15 or BK15+BAO data, compared to the theoretical predictions of selected inflationary models. This figure has been adapted from~\cite{Akrami:2018odb}. }
\label{Fig:Planck2018}
\end{figure}

Concerning the cosmological observables, and assuming the slow-roll approximation, we start by discussing the scalar and tensor power spectra which play a crucial role in inflationary cosmology. The CMB observations considerably restrict the inflationary predictions as it is shown in Fig~\ref{Fig:Planck2018}.
Choosing an arbitrary pivot scale $k_\star$ that exited the horizon, the scalar ($\mathcal{P}_\zeta $) and tensor ($\mathcal{P}_T$) power spectra can be written as
\begin{equation}
\mathcal{P}_\zeta (k)=A_s \left(\frac{k}{k_\star} \right)^{n_s -1}, \qquad  A_s\simeq\frac{1}{24\pi^2}\frac{V(\phi_\star)}{\epsilon_V(\phi_\star)}  \qquad \mbox{and} \qquad \mathcal{P}_T (k)\simeq\frac{2V(\phi_\star)}{3\pi^2} \left(\frac{k}{k_\star} \right)^{n_t}\,,
\end{equation}
where the scale-dependence of the power spectra is defined by the scalar ($n_s$) and tensor ($n_t$) spectral indices given by
\begin{equation}
n_s-1=\frac{{\rm d} \ln \mathcal{P}_\zeta (k) }{{\rm d} \ln k}= -6\epsilon_V +2\eta_V  \qquad \mbox{and} \qquad n_t= \frac{{\rm d} \ln \mathcal{P}_T (k) }{{\rm d} \ln k}\,.
\end{equation}
In these we have used the potential slow-roll parameters (note that in most formulas we use $M_{\rm P}^2 = 1$  except when we want the dimensionality to be explicit)
\begin{equation}
\epsilon_V=\frac{1}{2}\left(\frac{V'(\phi)}{V(\phi)} \right)^2 \qquad \mbox{and} \qquad \eta_V = \frac{V''(\phi)}{V(\phi)}\,.
\end{equation}
The tensor-to-scalar ratio is defined as
\begin{equation}
    r= \frac{\mathcal{P}_T (k)}{\mathcal{P}_\zeta (k)} = 16\epsilon_V.
\end{equation}
The Planck collaboration~\cite{Akrami:2018odb} has set the following bounds on the values of the observables:
\begin{equation}\label{eq:Planck_constr}
    A_s = (2.10\pm 0.03)\times10^{-9},\qquad n_s=0.9649 \pm 0.0042  \quad (1\sigma\mbox{ region}), \qquad r < 0.056\,.
\end{equation}
During inflation the variation of the scalar field is related to the so-called number of $e$-folds $N_\star$ which has to be between $50$ and $60$ for the horizon and flatness problems to be solved. Following~\cite{Akrami:2018odb,Liddle:2003as} the number of $e$-folds is given by
\begin{equation} 
    N_\star \simeq \int_{\phi_{\rm end}}^{\phi_\star} \frac{{\rm d} \phi}{\sqrt{2\epsilon_V}} = \ln \left[ \left(\frac{\pi^2}{30} \right)^{\frac{1}{4}} \frac{( g_{s,0})^{ \frac{1}{3}}}{ \sqrt{3} } \frac{T_0}{H_0} \right]
 - \ln\left[ \frac{k_\star}{  a_0 H_0} \right] 
 + \frac{1}{4} \ln\left[\frac{V^2(\phi_\star)}{\rho_{\rm end}} \right] - \frac{1 }{12}  \ln  \left[ g_{s,{ \rm reh}} \right]\,,
 \label{eq:efolds}
\end{equation}
where the subscripts $ ``0" $,  $``{\rm reh}"$ and $``{\rm end}"$ denote quantities at the present epoch, at the reheating phase and at the end of inflation respectively. The energy density is denoted by $\rho$. The entropy density degrees of freedom (DOF) have the values $g_{s,0}=43/11$ and $ g_{s,{ \rm reh}} = \mathcal{O}(100)$ for $T_{\rm reh}\sim 1\,\mathrm{TeV}$ or higher. At the present epoch the CMB temperature and the Hubble constant are $T_0=2.725\,\mathrm{K}$ and $H_0=67.6\,\mathrm{km/s/Mpc}$ respectively and we fix the pivot scale to $k_\star = 0.05\, \mbox{Mpc}^{-1}$ or $k_\star = 0.002\, \mbox{Mpc}^{-1}$.

Starobinsky~\cite{Starobinsky1980} and Higgs~\cite{Bezrukov2008} inflation are two of the simplest and most successful models of inflation. In the Starobinsky model a $\mathcal{R}^2$ is added in the usual Einstein-Hilbert action which after a Weyl rescaling of the metric and a field redefinition is converted to a scalar DOF. If initially a scalar field is contained in the action, then two-field inflation takes place. In Higgs inflation a nonminimal coupling between gravity and the Higgs field $h$ is added in the action. The actions in the Jordan frame (JF) read 
\begin{eqnarray}
    S_{JF}^{\rm S} &=& \int \mathrm{d}^4x\sqrt{-g} \left(\frac{M^2_{\rm Pl} \mathcal{R}}{2} + \frac{\mathcal{R}^2}{12 M^2}\right)\,, 
    \\  S_{JF}^{\rm H} &=& \int \mathrm{d}^4x\sqrt{-g} \left( \frac{(M^2_{\rm Pl}+\xi_h h^2) \mathcal{R}}{2} -\frac{1}{2}(\partial h)^2  -\frac{\lambda_h}{4} (h^2-v_h^2)^2\right)\,,
\end{eqnarray}
while the corresponding EF inflationary potentials for the canonical normalized scalar fields are
\begin{equation}
    \bar{U}^{\rm S}(\phi) = \frac{3 M^2 M^4_{\rm Pl}}{4} \left[ 1- \exp\left(-\sqrt{\frac{2}{3}} \frac{\phi}{M_{\rm Pl}} \right)\right]^2\,,\,
    \bar{U}^{\rm H}(\phi) =  \frac{\lambda_h M^4_{\rm Pl}}{4 \xi_h^{2}}\left(1+\exp \left(-\sqrt{\frac{2}{3}} \frac{\phi}{M_{\rm Pl}} \right)\right)^{-2}\,.
\end{equation}
The parameters $M$ and $\xi_h$ are determined by the constraint on the amplitude of the scalar power spectrum. The rest cosmological observables are well within the allowed region in Fig.~\ref{Fig:Planck2018} as for both models $n_s=1-2/N_\star$ and $r=12/N_\star^2$.

\section{Palatini formulation of gravity}

In the context of Palatini gravity~\cite{Palatini1919,Ferraris1982}, which is an interesting alternative to the usual metric theory
of gravity, the metric tensor and the connection are treated as independent variables and one has to vary the action with respect to metric and the connection. In Palatini formulation the addition of higher order in curvature terms will not generate a propagating scalar DOF in the EF as in the metric case~\cite{Kaiser2014,Kaneda:2015jma,Myrzakulov:2015qaa,Odintsov2016,Ema:2017rqn,Mori:2017caa,He:2018gyf,Gundhi:2018wyz,Elizalde:2018now,He:2018mgb,Canko:2019mud}. The differences between the formulations in the cosmological predictions, as far as inflation is concerned, arise from the nonminimal couplings of the scalars~\cite{Bauer:2008zj,Bauer:2010bu,Tamanini:2010uq,Bauer:2010jg,Rasanen:2017ivk,Tenkanen:2017jih,Racioppi:2017spw,Markkanen:2017tun,Jarv:2017azx,Fu:2017iqg,Racioppi:2018zoy,Carrilho:2018ffi,Kozak:2018vlp,Rasanen:2018fom,Rasanen:2018ihz,Almeida:2018oid,Shimada:2018lnm,Takahashi:2018brt,Jinno:2018jei,Rubio:2019ypq,Bostan:2019uvv,Bostan:2019wsd,Tenkanen:2019xzn,Racioppi:2019jsp,Tenkanen:2020dge,Borowiec:2020lfx,Jarv:2020qqm,Karam:2020rpa,McDonald:2020lpz,Gialamas:2020vto,Verner:2020gfa,Enckell:2020lvn,Reyimuaji:2020goi,Karam:2021wzz,Mikura:2021ldx,Kubota:2020ehu,Saez-ChillonGomez:2021byq} or from higher order in curvature terms~\cite{Olmo:2011uz,Bombacigno:2018tyw,Enckell:2018hmo,Antoniadis:2018ywb,Antoniadis:2018yfq,Tenkanen:2019jiq,Edery:2019txq,Giovannini:2019mgk,Tenkanen:2019wsd,Gialamas:2019nly,Antoniadis:2019jnz,Tenkanen:2020cvw,Lloyd-Stubbs:2020pvx,Antoniadis:2020dfq,Ghilencea:2020piz,Das:2020kff,Gialamas:2020snr,Ghilencea:2020rxc,Bekov:2020dww,Dimopoulos:2020pas,Gomez:2020rnq,Karam:2021sno,Lykkas:2021vax,Gialamas:2021enw} and has attracted the interest of many authors. 
Theories containing scale invariant quadratic terms in the context of metric formulation has recently received a lot of attention as a possible realization of quantum gravity~\cite{Stelle:1976gc,Biswas:2005qr,Salvio2014,Edery:2014nha,Farzinnia:2015fka,Salvio2018,Salvio:2018crh,Edery:2019bsh,Ghilencea:2019rqj,Salvio:2019llz}. We will consider similar theories but in the Palatini case.

Starting from the JF Palatini action
\begin{equation}
\label{eq:JF_action}
S_{JF} = \int \mathrm{d}^4x\sqrt{-g} \left(\xi_\phi\phi^2 \frac{R}{2}+\frac{\alpha}{2}R^2 +\frac{\beta}{2}R_{(\mu\nu)}R^{(\mu\nu)} -\frac{1}{2}(\partial \phi)^2-V(\phi) \right)
\end{equation}
and after a Weyl rescaling of the form $g_{\mu\nu} \rightarrow \Omega^2 g_{\mu\nu}$ with $\Omega^2 = \xi_\phi \phi^2$, we can pass to an intermediate frame (IF)
\begin{equation}
\label{eq:IF_action}
S_{IF} = \int \mathrm{d}^4x\sqrt{-g} \left( \frac{R}{2}+\frac{\alpha}{2}R^2 +\frac{\beta}{2}R_{(\mu\nu)}R^{(\mu\nu)} -\frac{1}{2\Omega^2}(\partial \phi)^2-U(\phi) \right)\,,
\end{equation}
with $U(\phi)=V(\phi)/\Omega^4$. Note also that the Planck scale is dynamically generated via the vacuum expectation value (VEV) of the scalar field as $M^2_{\rm Pl} = \xi_\phi v_\phi^2$. This is usually achieved in scale-invariant theories~\cite{Shaposhnikov2009a,GarciaBellido:2011de,Khoze2013,Steele:2013fka,Ren:2014sya,Kannike2014,Csaki2014,Kannike2015a,Kannike2016b,Wang:2015cda,Farzinnia:2015fka,Rinaldi2016a,Marzola2016b,Ferreira2016,Kannike2017a,Marzola2016,Kannike2017,Ferreira:2016wem,Karam2017,Karam2017a,Racioppi2018,Ferreira:2018a,Wetterich:2019qzx,Vicentini:2019etr,Ghilencea:2019rqj,Salvio:2019wcp,Racioppi:2019jsp,Benisty:2020nuu,Tang:2020ovf,Gialamas:2020snr,Ghilencea:2020rxc,Gialamas:2021enw}, where the running of the inflaton quartic coupling induces symmetry breaking \`{a} la Coleman--Weinberg. Now, including an auxiliary field and after a conformal (or a disformal transformation~\cite{Bekenstein:1992pj,Zumalacarregui:2013pma,Afonso:2017bxr,Annala2020} in the $\beta \neq 0$ case), we obtain the final EF action 
\begin{equation}
\label{eq:EF_action}
S_{EF} = \int \mathrm{d}^4x\sqrt{-g} \left( \frac{R}{2} -\frac{1}{2(1+\tilde{\alpha}U(s_c))} (\partial s_c)^2-\bar{U}(s_c) +\mathcal{O}\left( (\partial s_c)^4 \right)\right)\,,
\end{equation}
where $\tilde{\alpha}=2\beta+8\alpha$ and $s_c$ is the canonical field in the IF. The EF inflationary potential reads
\begin{equation}
\label{eq:pot_EF}
\bar{U}(s_c)=U(s_c)/(1+\tilde{\alpha}U(s_c))\,.
\end{equation}
The potential $\bar{U}(s_c)$ reaches a plateau $1/\tilde{\alpha}$ for large field values. This flatness is quite capable to reduce the tensor-to-scalar ratio.

\section{Nonminimal Coleman-Weinberg inflation in Palatini quadratic gravity}

In~\cite{Gialamas:2020snr} we considered the nonmimimal Coleman-Weinberg (CW) model~\cite{Coleman1973} with a $R^2$ term $(\alpha \neq 0\,,\, \beta=0)$ in the Palatini formulation. The scalar potential contains a running quartic coupling $\lambda(\phi)$ and a cosmological constant $\Lambda$, that is
\begin{equation}
V(\phi) = \frac{1}{4} \lambda (\phi) \phi^4 + \Lambda^4.
\label{eq:CW_pot}
\end{equation}
At the minimum we require that $V(v_\phi)= 0$, with $v^2_\phi= M^2_{\rm Pl}/\xi_\phi$. The minimization condition $\beta(v_\phi) +4\lambda(v_\phi)=0$ implies that, a) $\beta (v_\phi) > 0, \ \lambda (v_\phi)<0$ or b) $\beta (v_\phi) =\lambda (v_\phi)=0$, where $\beta (v_\phi)$ is the beta-function of the quartic coupling. Expanding the quartic coupling around the VEV and using the cases a) and b) we get that
\begin{figure}[t]
\centering
\includegraphics[width=0.49\linewidth]{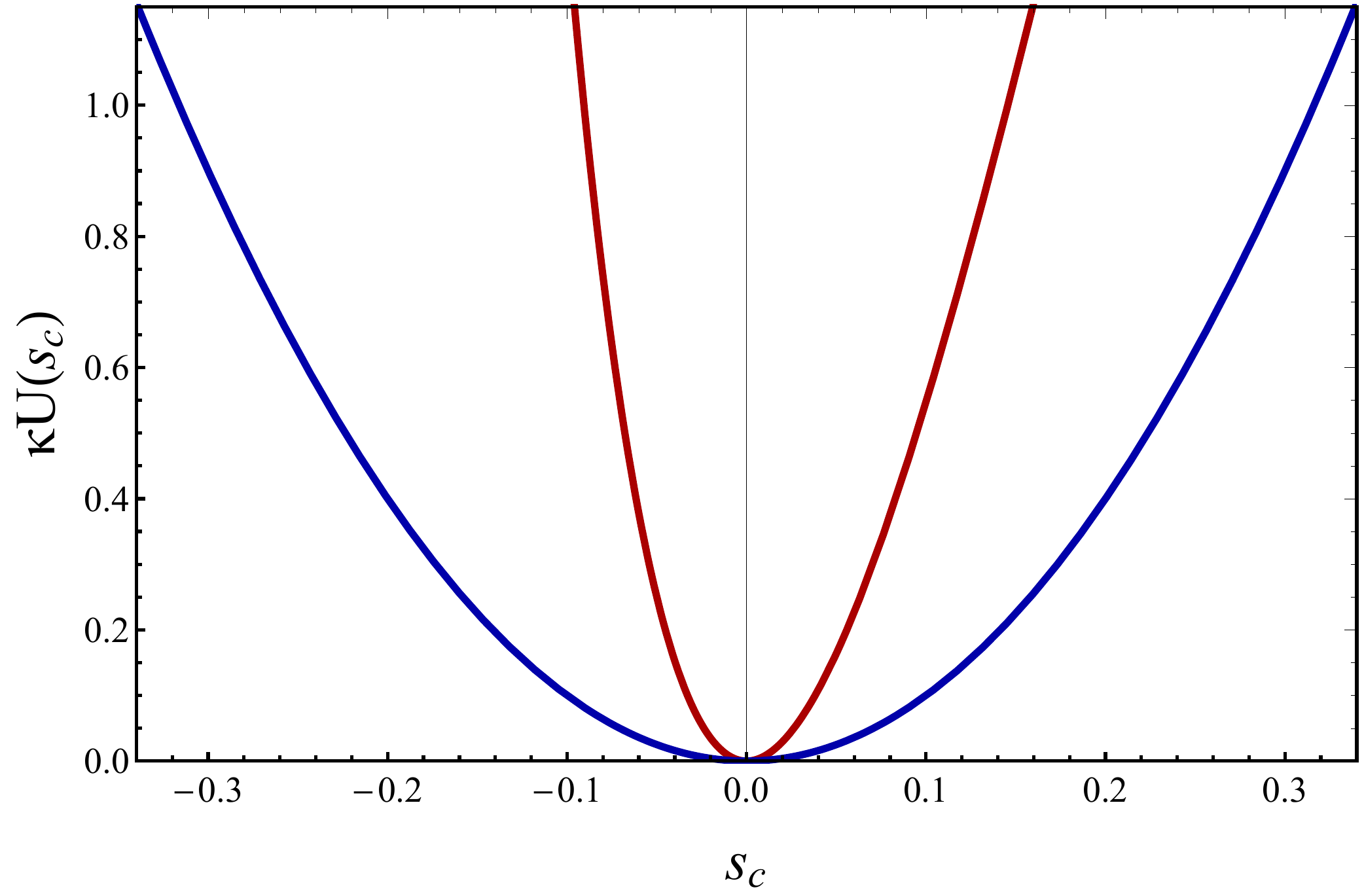}
\includegraphics[width=0.49\linewidth]{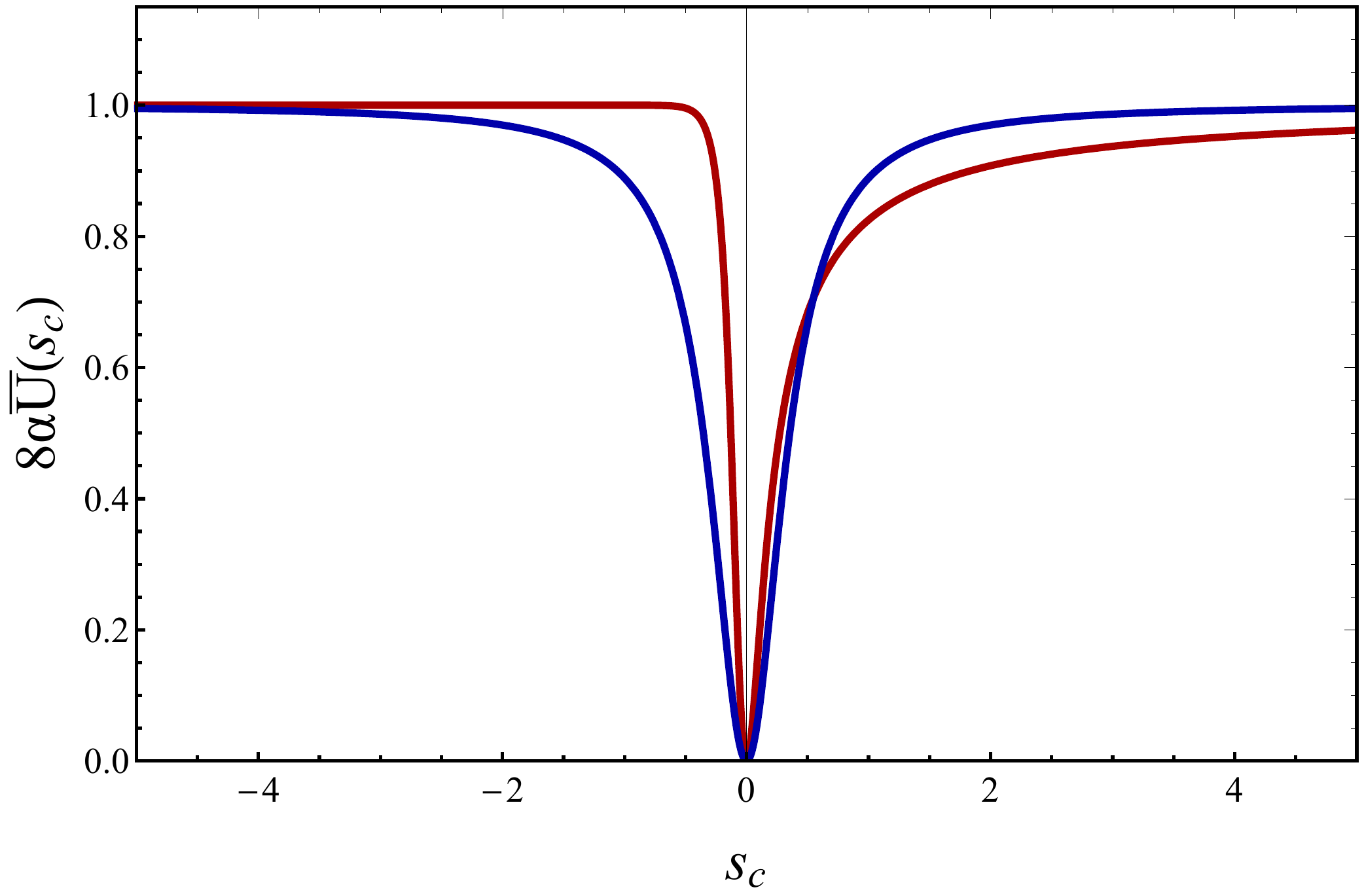}
\caption{\textbf{Left:}  Scalar potential $U(s_c)$ of Eq.(\ref{eq:CW_pot_IF}) for the 1st order CW model  (red line) with $\kappa = \Lambda^{-4}$ and for the 2nd order CW model (blue line) with $\kappa = \xi_\phi^2/ \beta'$. \textbf{Right:} Scalar potential $\bar{U}(s_c)$ of Eq.(\ref{eq:pot_EF}) for the 1st order CW model  (red line) and for the 2nd order CW model  (blue line).}
\label{Fig:CW_pots}
\end{figure}
\begin{equation} 
\lambda^a(\phi) \simeq \lambda(v_\phi) + \beta(v_\phi) \ln\frac{\phi}{v_\phi} \, , \qquad  \lambda^b(\phi) \simeq  \frac{ \beta'(v_\phi)}{2} \ln^2\frac{\phi}{v_\phi}\,,
\end{equation}
which after substituting in~(\ref{eq:CW_pot}) give the 1st and 2nd order JF CW potentials
\begin{equation}
V_1(\phi)= \Lambda ^4 \left\{ 1 + \left[ 4 \ln \left(\frac{\phi}{v_\phi}\right) -1 \right] \frac{\phi^4}{v_\phi^4} \right\} \, , \qquad V_2(\phi) = \frac{1}{8} \beta'  \phi ^4 \ln ^2\left(\frac{\phi }{v_\phi}\right) \,.
\label{eq:CW_pot_JF}
\end{equation}
In the canonical normalized IF~(\ref{eq:IF_action}) the potentials~(\ref{eq:CW_pot_JF}) are of the form 
\begin{equation}
U_1(s_c)=\Lambda ^4 \left(4 \sqrt{\xi_\phi} \frac{s_c}{ M_{\rm Pl}}+e^{-4  \sqrt{\xi_\phi} \frac{s_c}{ M_{\rm Pl}}}-1\right) \, , \qquad U_2(s_c)= \frac{ \beta' M^2_{\rm Pl}}{8\xi_\phi} s_c^2\,.
\label{eq:CW_pot_IF}
\end{equation}
The 1st order potential behaves like a linear potential $U_1(s_c) \propto s_c$ for $\xi_\phi \gg 1$ (and $s_c >0$) and as a quadratic potential $U_1(s_c) \propto s_c^2$ for $\xi_\phi \ll 1$. The 2nd order is completely quadratic in the IF.
Therefore the inflationary potential in the EF~(\ref{eq:EF_action}) after the effect of the $R^2$ term is given by~(\ref{eq:pot_EF}) with $\tilde{\alpha}=8\alpha$. In Fig.~\ref{Fig:CW_pots} are displayed the IF potentials~(\ref{eq:CW_pot_IF}) (left) and the EF potentials~(\ref{eq:pot_EF}) (right) for $\xi_\phi=10$, $\alpha=10^{10}$, $\beta'= 10^{-9}$ and $\Lambda=0.0015$. The normalized factor is $\kappa = \Lambda^{-4}$ or $k=\xi_\phi^2/\beta'$ depending on the case. As is obvious the final EF inflationary potentials in the right panel are flat for large field values, approaching the plateau $1/8\alpha$. 

\begin{figure}[t]
\centering
\includegraphics[width=0.49\linewidth]{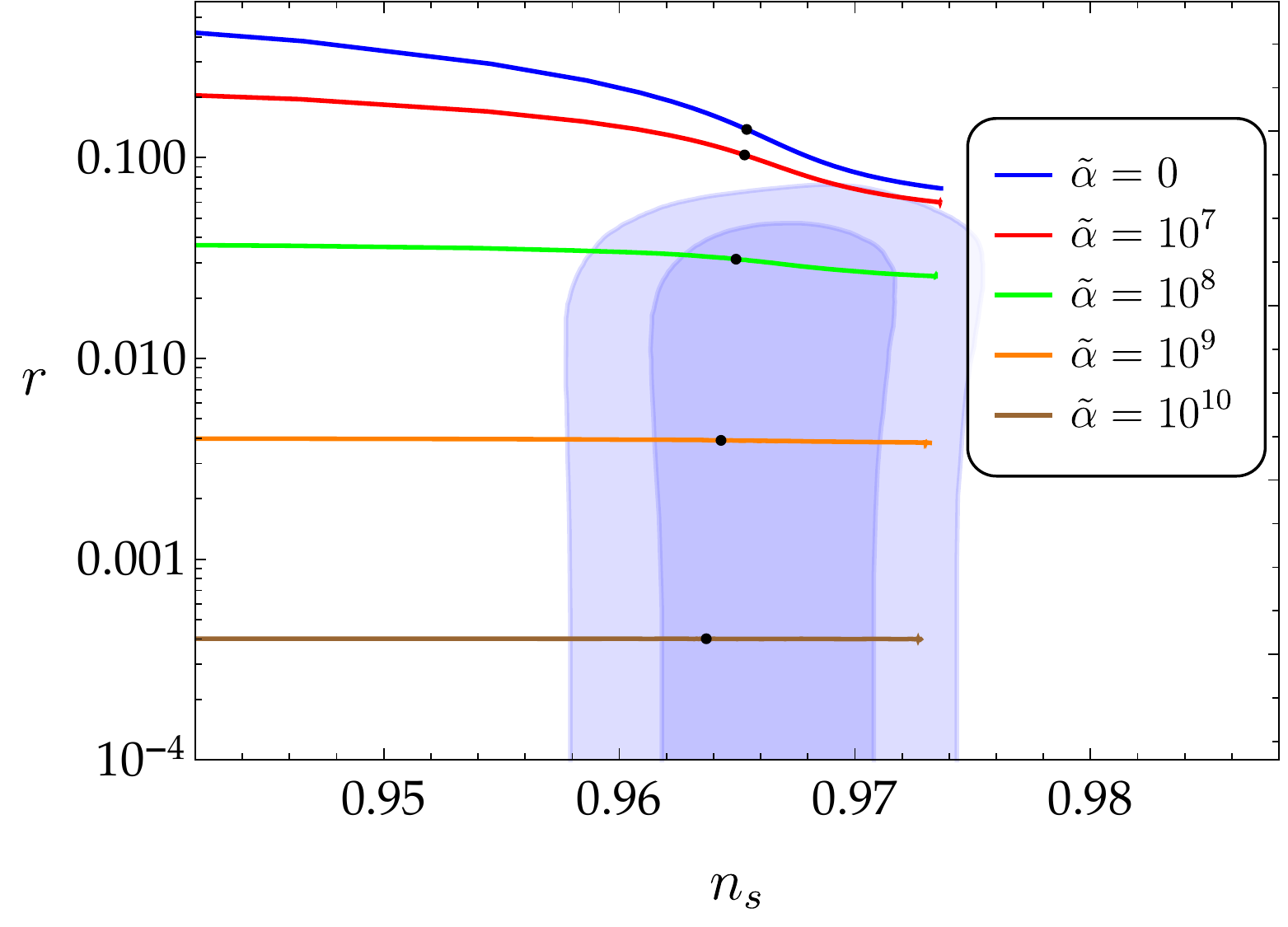}
\includegraphics[width=0.49\linewidth]{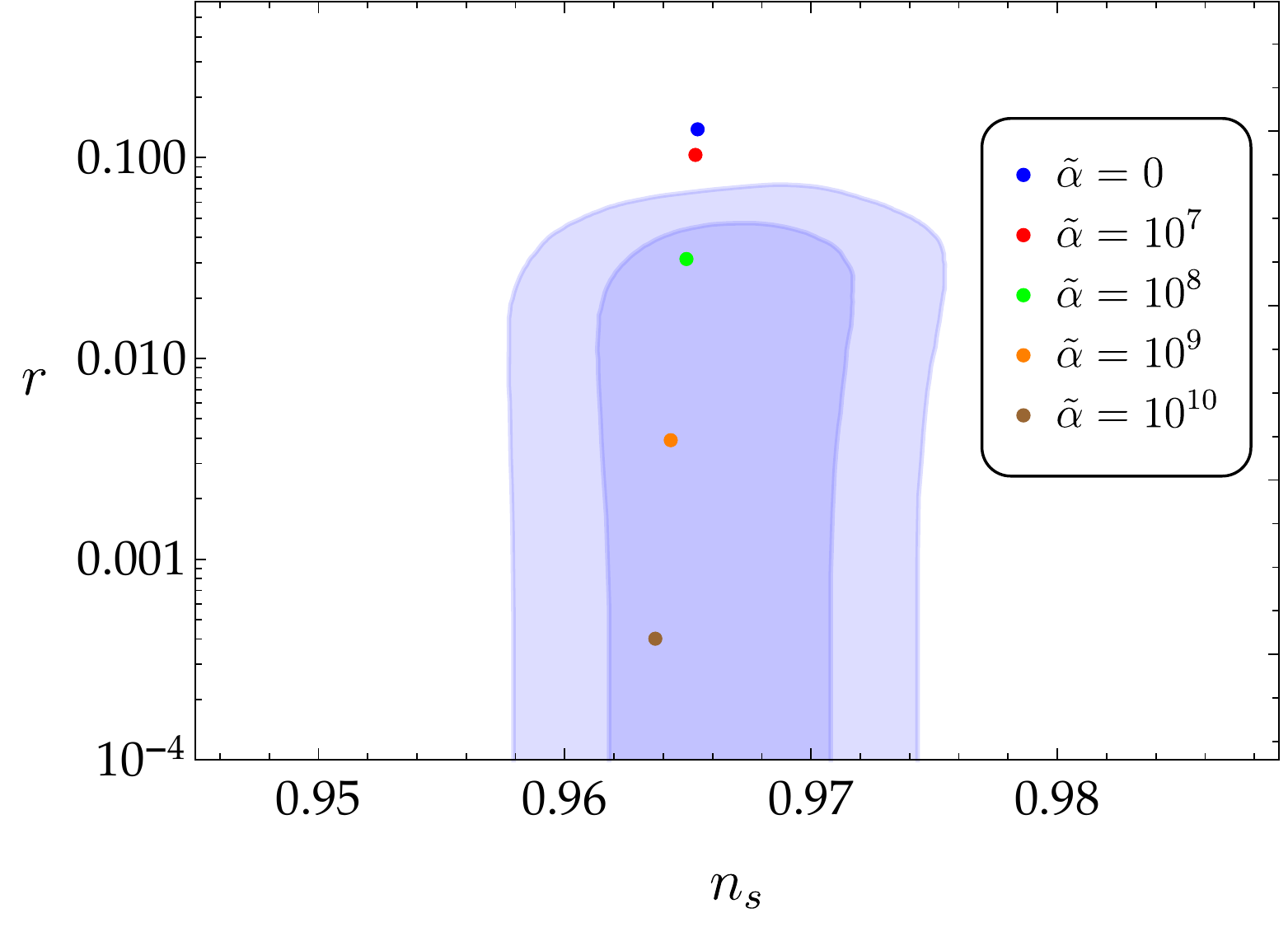}
\caption{\textbf{Left:} The predictions for the tensor-to-scalar ratio ($r$) and the spectral index ($n_{\rm s}$)  for the 1st order CW potential. \textbf{Right:} The predictions for the tensor-to-scalar ratio ($r$) and the spectral index ($n_{\rm s}$)  for the 2nd order CW potential. }
\label{Fig:CW_obs}
\end{figure}

Finally, in Fig.~\ref{Fig:CW_obs} we present the inflationary predictions for the 1st (left) and 2nd (right) order potentials, for various values of the parameter $\alpha$.  For each curve, in the 1st order potential, the black dot corresponds to the $\xi_\phi \rightarrow 0$ limit. The parts of the curves in the left of the black dots correspond to small field inflation (SFI), while the right ones to large field inflation (LFI).  The number of $e$-folds are predefined by Eq.~(\ref{eq:efolds}) for $k_\star = 0.05\, \mbox{Mpc}^{-1}$.

\section{Scale invariant inflation with a $U(1)_X$ extension of the Standard Model in Palatini quadratic gravity}
In~\cite{Gialamas:2021enw} we considered a $U(1)_X$ extension of the Standard Model (SM)~\cite{Hempfling1996,Chang:2007ki,Iso2009,Englert2013,Khoze2013a,Benic:2014aga,Das:2015nwk,Marzo:2018nov} in which the extra particles are three right-handed (RH) neutrinos $N_R$, one gauge boson and one scalar field $\Phi$, which in the unitary gauge is given by $\Phi =(\phi+v_\phi)/\sqrt{2}$. The relevant Lagrangians of the model are
\begin{eqnarray}
\bullet &&\mathcal{L}^{\rm BSM}_{\rm Yukawa} = -y_D^{i j} \overline{\ell}^{i}_{L} H N^{j}_{R} -\frac{1}{2} y_M^{i} \Phi  \overline{N}^{i C}_{R} N^{i}_{R} + h.c.\,,\nonumber \\
\bullet  &&\mathcal{L}^{\rm BSM}_{\rm scalar} =  -\frac{1}{2} g^{\mu\nu} \partial_\mu \phi \partial_\nu \phi - \frac{1}{4} \lambda_\phi \phi^4 +\frac{1}{4} \lambda_{h \phi} h^2\phi^2\,, \nonumber \\
\bullet &&\mathcal{L}_{\rm SM} \qquad \mbox{with no Higgs mass term}\,, \nonumber \\
\bullet  &&\mathcal{L}_{\rm gravity} =  \frac{1}{2} \left( \xi_\phi \phi^2 + \xi_h h^2 \right) g^{\mu\nu} R_{\mu\nu} + \frac{\alpha}{2}R^2 +\frac{\beta}{2} R_{(\mu\nu)}R^{(\mu\nu)}\,.
\end{eqnarray}
The mass of the Higgs and the electroweak scale is generated through a portal coupling of the form $\lambda_{h \phi} h^2 \phi^2$. Thus, the addition of the extra scalar field $\phi$ is necessary to preserve the scale invariance of our model since the known Higgs mass term contained in the SM Lagrangian is not scale invariant. Also, the Planck scale is dynamically generated when $\phi$ and $h$ develop their VEVs, that is $M^2_{\rm P} = \xi_\phi v_\phi^2 + \xi_h v_h^2$.

The JF tree-level potential
\begin{equation}
\label{eq:scalar_pot}
 V^{(0)}(\phi,h) =\frac{1}{4}\left( \lambda_\phi \phi^4- \lambda_{h\phi} h^2 \phi^2 + \lambda_{h} h^4 \right)\,,
\end{equation}
will be studied with the help of the Gildener-Weinberg formalism~\cite{Gildener:1976ih}. In this approach one first finds the flat direction (FD) of the tree-level potential and then computes the one-loop corrections only along this flat direction, since this is where the corrections play the dominant role. For reasons that will be clear later we will calculate the FD of the IF potential $U^{(0)}(\phi,h) =V^{(0)}(\phi,h)/\left( \xi_\phi \phi ^2 + \xi_h h ^2 \right)^2$. The FD is determined by the conditions $\partial_{\phi} U^{(0)}(\phi,h) = \partial_{h} U^{(0)}(\phi,h) = 0$, which give that along the FD
\begin{figure}[t]
\centering
\includegraphics[width=0.5\linewidth]{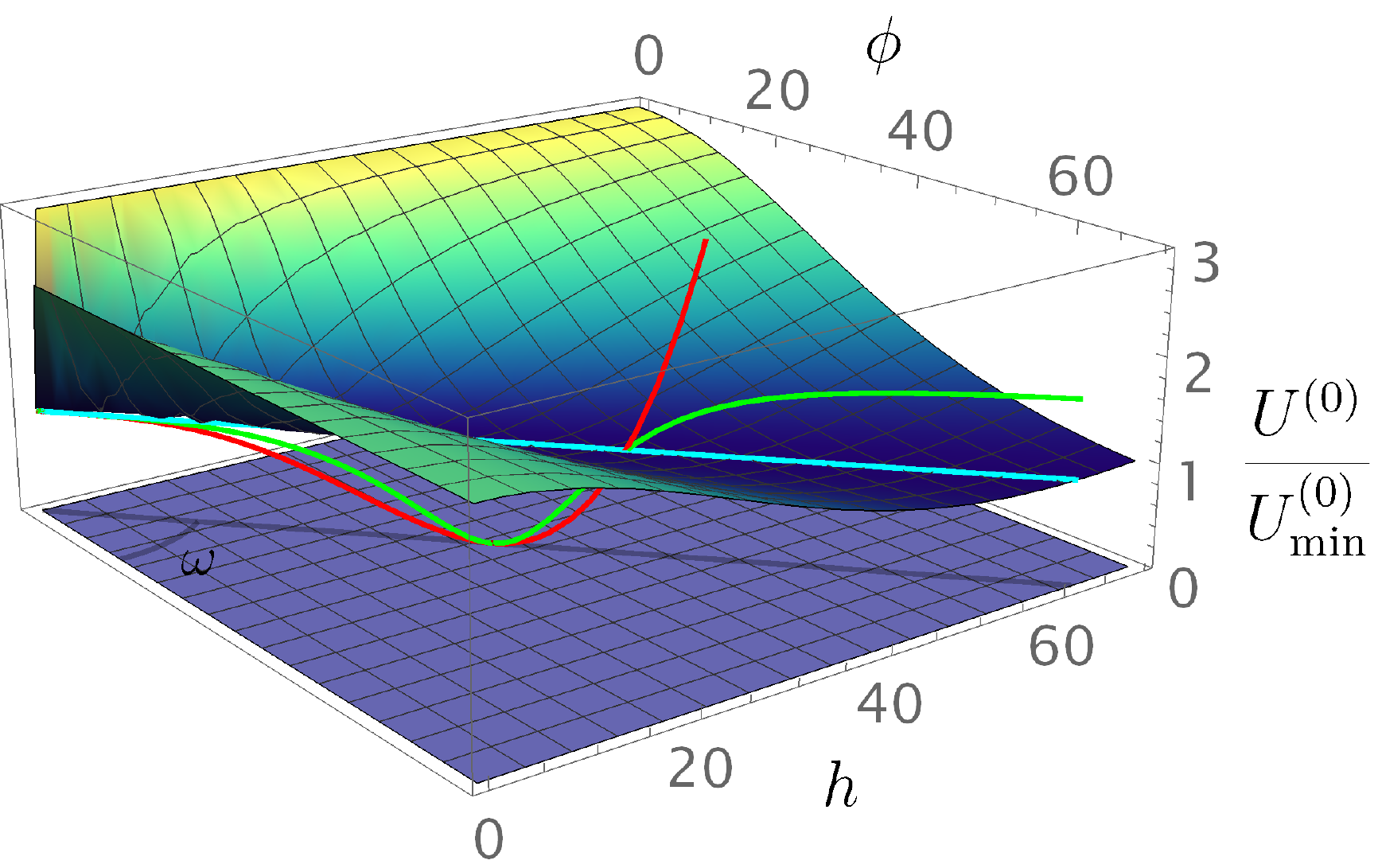}
\includegraphics[width=0.49\linewidth]{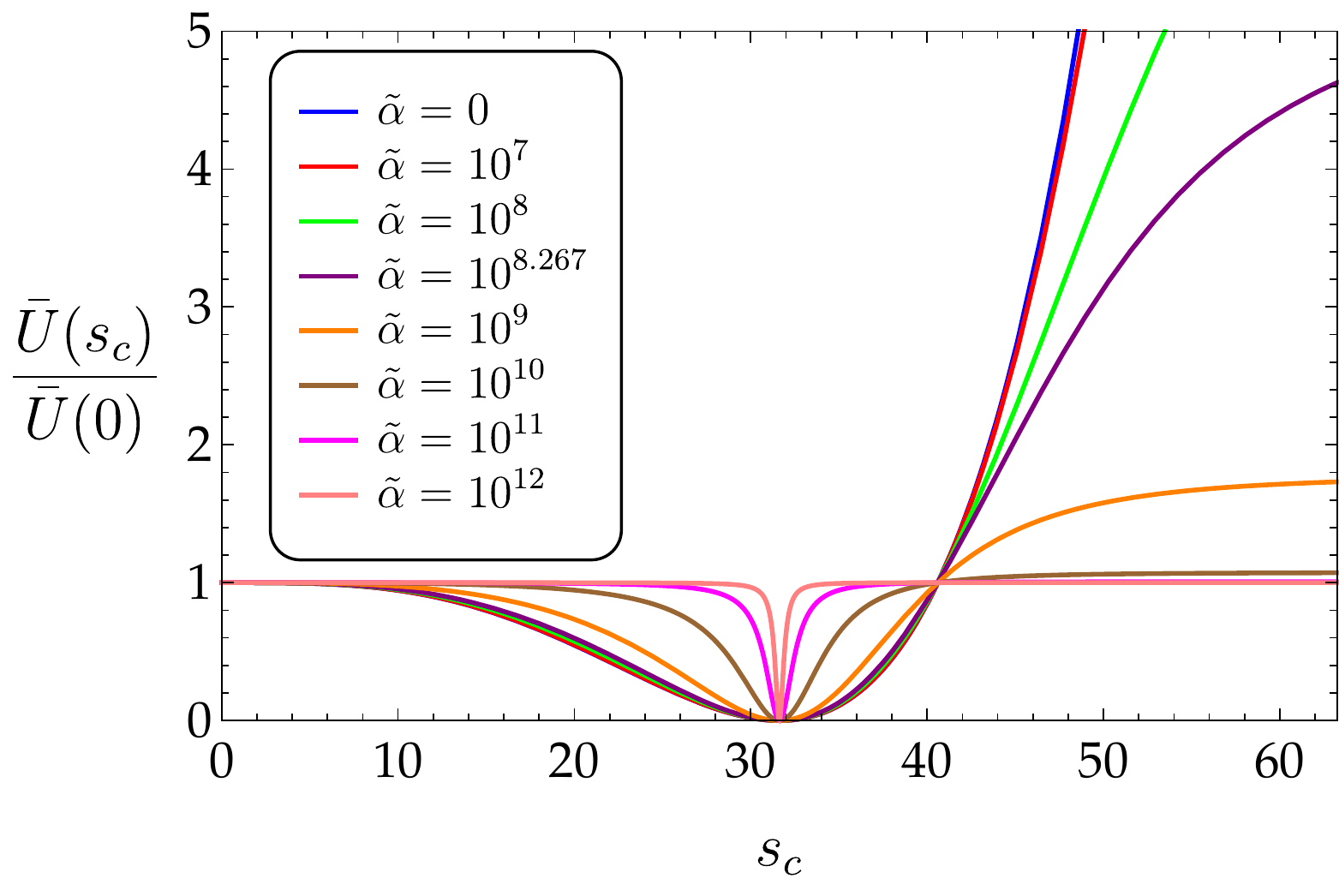}
\caption{\textbf{Left:} The normalized tree-level potential $U^{(0)}(\phi,h)/U^{(0)}_{\rm min}$ and its flat direction (cyan line). The red curve corresponds to the normalized one-loop corrected IF potential along the flat direction $U_{\rm eff} (s_c)/U_{\rm eff} (0)$, while the green one to the normalized EF potential $\bar{U}(s_c)/\bar{U}(0)$. The values of the parameters are $\tilde{\alpha} =10^9 ,\,\xi_s=0.001 ,\,$ and $\mathcal{M} \simeq 0.0357$. For illustrative purposes we have chosen the unrealistic value for the mixing angle $\omega \simeq 0.732$. \textbf{Right:} The normalized inflationary EF potential $\bar{U}(s_c)$ for same $\xi_s\,,\,\mathcal{M}$ and various values of the parameter $\tilde{\alpha}$. }
\label{Fig:pots_3d}
\end{figure}
\begin{equation}
U^{(0)}_{\rm min} \equiv U^{(0)}(v_\phi,v_h)=\frac{\left ( 4 \lambda_h \lambda_\phi - \lambda_{h \phi}^2 \right) M^4_{\rm P}}{16 \left[ \lambda_\phi \xi_h^2 + \xi_\phi \left ( \lambda_{h \phi} \xi_h+\lambda_h \xi_\phi \right ) \right]}\,.
\label{eq:U0min}
\end{equation}
In order to move from the initial frame of fields $\left(\phi,h\right)$ to the FD frame $\left(s,\sigma \right)$, where the direction of the so-called scalon field $s$ is identified with the FD and $\sigma$ is the perpendicular direction, we perform an orthogonal rotation described by the transformation
\begin{equation}
\left(\begin{array}{c}
\phi \\
h 
\end{array}
\right) = \begin{pmatrix} 
\cos{\omega} & -\sin{\omega} \\
\sin{\omega} & \cos{\omega}
\end{pmatrix} \left(\begin{array}{c}
s \\
\sigma
\end{array}
\right) \,,
\end{equation}
where the mixing angle is given by $\omega = \arctan{\left( v_h/v_\phi \right)}$. Along the FD ($\sigma =0$) the only relevant DOF is the scalon $s$ which is related to $\phi$ and $h$ via
$s^2 = \phi^2 + h^2$. As a consequence an effective nonminimal coupling constant is defined as $\xi_s = \xi_\phi \cos^2 \omega + \xi_h \sin^2 \omega$. Finally, we perform the following field redefinition in order to render the kinetic term of $s$ canonical:
\begin{equation}
s_c - v_c = \int_{v_s}^s \frac{1}{\sqrt{\xi_s}} \frac{{\rm d} s'}{s'} = \frac{1}{\sqrt{\xi_s}} \ln \frac{s}{v_s}\,.
\end{equation}
The VEV $v_s$ of the scalon is related to the Planck mass with the familiar equation $v^2_s = M^2_{\rm P}/\xi_s$.

The one-loop corrections along the flat direction for the canonical field $s_c$ at the scale $\Lambda$ may be written as
\begin{equation}
\label{eq:one-loop-cor}
U^{(1)} (s_c) = \mathbb{A}\, s_c^4 + \mathbb{B}\, s_c^4 \ln \frac{s_c^2}{\Lambda^2} \,,
\end{equation}
where in our model
\begin{eqnarray}
\mathbb{A} &=& \frac{1}{64 \pi^2 v_s^4} \left\lbrace M_h^4 \left( \ln \frac{M_h^2}{v^2_s} - \frac{3}{2} \right) + 6 M_W^4 \left( \ln \frac{M_W^2}{v^2_s} - \frac{5}{6} \right) + 3 M_Z^4 \left( \ln \frac{M_Z^2}{v^2_s} - \frac{5}{6} \right)  \right. \nonumber \\
 && \left. \qquad \ + 3 M_X^4 \left( \ln \frac{M_X^2}{v^2_s} - \frac{5}{6} \right) - 6 M_{N_R}^4 \left( \ln \frac{M_{N_R}^2}{v^2_s} - 1 \right) - 12 M_t^4 \left( \ln \frac{M_t^2}{v^2_s} - 1 \right)  \right\rbrace\,, \\ 
\mathbb{B} &=& \frac{\mathcal{M}^4}{64\pi^2 v^4_s} \,, \qquad \mathcal{M}^4 \equiv M_h^4 + 3 M_X^4 + 6 M_W^4 + 3 M_Z^4 - 6 M_{N_R}^4 - 12 M_t^4\,.
\end{eqnarray}
Minimizing~\eqref{eq:one-loop-cor}, we can determine the scale $\Lambda$ as $\Lambda = v_s \exp \left[ \frac{\mathbb{A}}{2 \mathbb{B}} + \frac{1}{4} \right]$. Then, the one-loop correction can be expressed as
\begin{equation}
U^{(1)} (s_c) = \frac{\mathcal{M}^4}{64\pi^2 v^4_s} s_c^4 \left[ \ln \frac{s_c^2}{v^2_s}  - \frac{1}{2} \right] \,.
\label{eq:one_loop}
\end{equation}
In order the potential~(\ref{eq:one_loop}) to be bounded from below, the parameter $\mathcal{M}^4$ must be positive. This is achieved if $3 M_X^4 - 6 M_{N_R}^4 \gtrsim \left( 317\ \rm GeV \right)^4$, where $M_X$ is the mass of the extra $U(1)_X$ gauge boson and $M_{N_R}$ are the RH neutrinos masses. 
\begin{figure}[t]
\centering
\includegraphics[width=0.5\linewidth]{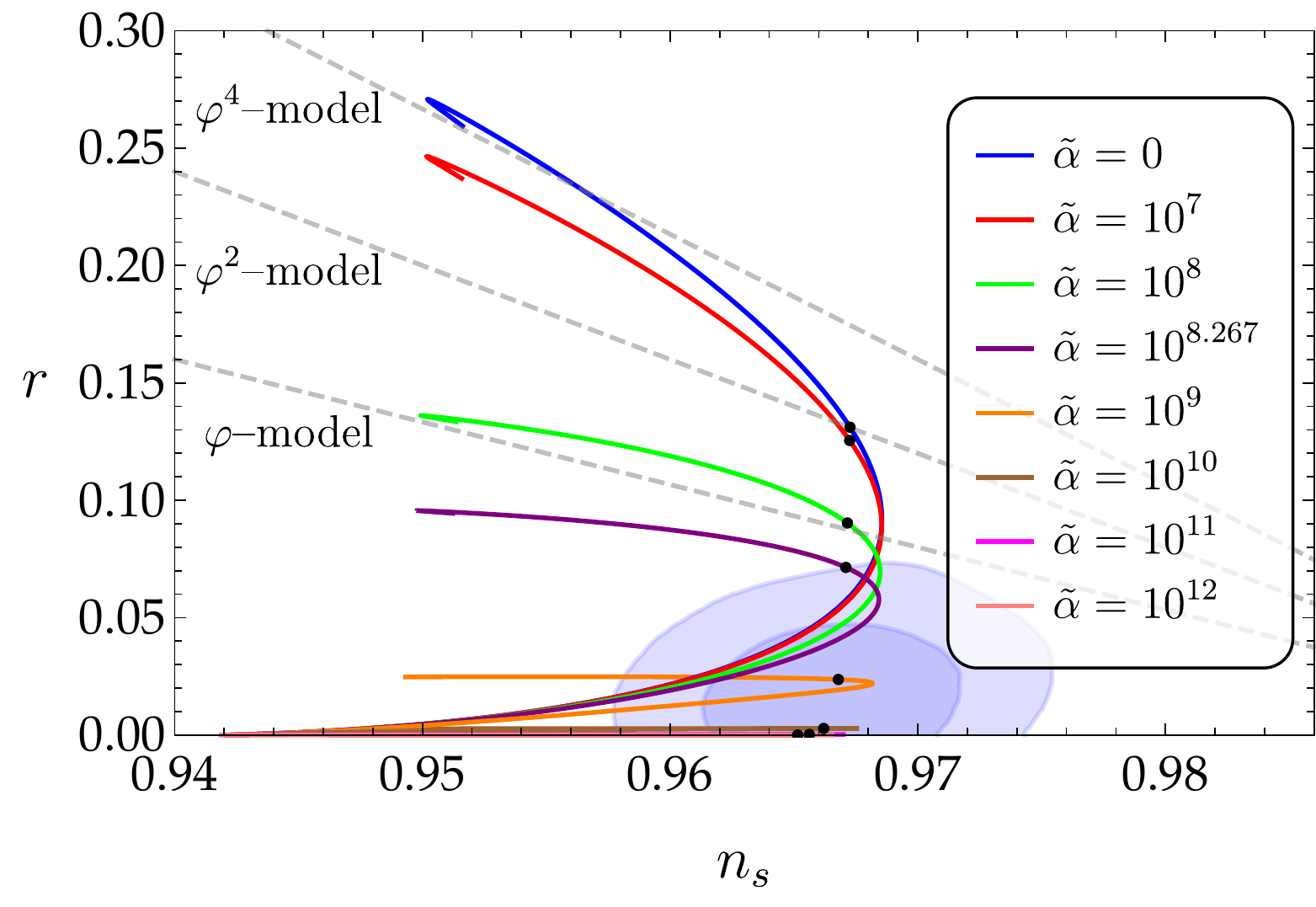}
\includegraphics[width=0.49\linewidth]{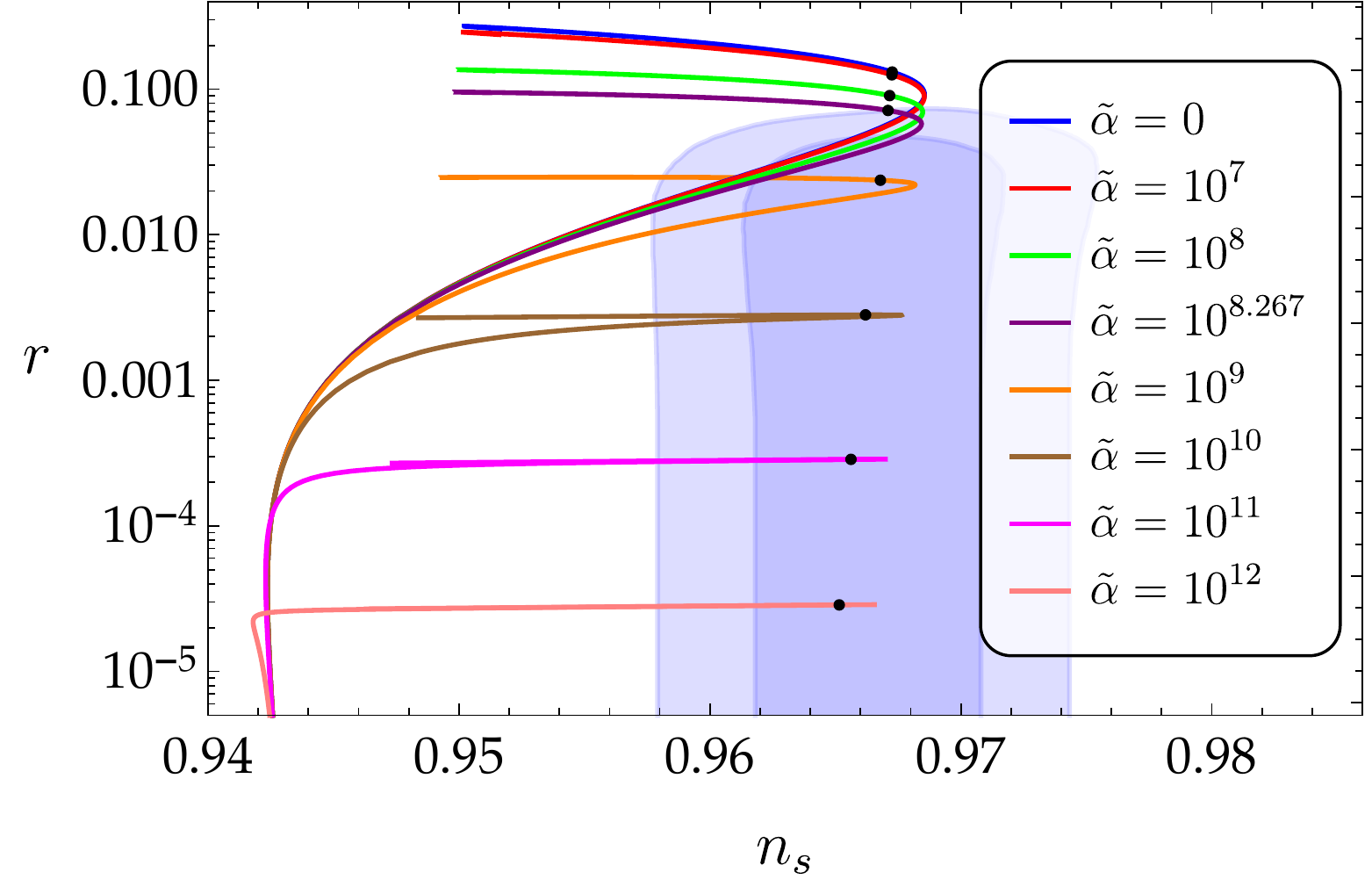}
\caption{ \textbf{Left:} The predictions for the tensor-to-scalar ratio ($r$) and the spectral index ($n_{\rm s}$) as $\xi_s$ ranges from $\xi_s \ll 1$ to $\xi_s \gg 1$ for various values of $\tilde{\alpha}$.  For each curve, the black dot corresponds to the $\xi_s \rightarrow 0$ limit. \textbf{Right:} The same as the left, but in logarithmic scale.}
\label{Fig:u1_pred}
\end{figure}

Requiring that the full one-loop effective potential is zero at $v_s$, i.e. $U_{\rm eff}(v_s) = U^{(0)}_{\rm min} + U^{(1)} (v_s) = 0$ we obtain that
\begin{equation}
U_{\rm eff} (s_c) = \frac{\mathcal{M}^4}{128 \pi^2} \left[ \frac{s^4_c}{v^4_s} \left( 2 \ln \frac{s^2_c}{v^2_s} - 1 \right)  + 1  \right]\,.
\end{equation}
Note that we assume that $ 4 \lambda_h \lambda_\phi - \lambda_{h \phi}^2>0$, so that $U^{(0)}_{\rm min}>0$. 
Had we opted to consider the one-loop corrections in the JF, and not in the IF, the extremization conditions for the tree-level JF potential would fix its value to zero along the flat direction and in conjunction with the fact that at the minimum the one-loop correction~\eqref{eq:one_loop} is negative,  the full one-loop effective potential  would correspond to an anti-de Sitter vacuum.

Figure~\ref{Fig:pots_3d} illustrates the various potentials in the different frames of the model. As it is shown in the right panel, in the limit $\tilde{\alpha} \gg 1$ the potential becomes symmetric around its VEV and consequently the predictions for LFI and SFI are identical. This is also visible from Fig.~\ref{Fig:u1_pred}, as the upper parts of the curves (LFI) coincide to the lower parts (SFI) as the parameter $\tilde{\alpha}$ becomes larger. In this figure the pivot scale $k_\star = 0.002\, \mbox{Mpc}^{-1}$ has been used in order to determine the number of $e$-folds from Eq.~(\ref{eq:efolds}), but the amplitude of the scalar power spectrum is fixed to $2.1\times 10^{-9}$ at the scale $k_\star = 0.05\, \mbox{Mpc}^{-1}$. Finally, we have seen that for viable inflation $\xi_s \lesssim 4\times 10^{-3} $, which in turn implies a lower cutoff, $v_s  \gtrsim  15\, M_{\rm P}$.

\section{Conclusions}
In conclusion, in the Palatini formulation, the higher order in curvature terms provide us an 
effective potential which is asymptotically flat for large field values, and as a consequence the tensor-to-scalar ratio can be significantly reduced. The Planck scale can be dynamically generated through the VEVs of the scalar fields, because of the terms $\xi_i \phi_i^2 R$ containing in the action. The Higgs mass and the electroweak scale is generated through a possible portal coupling of the form $\lambda_{h \phi} h^2 \phi^2$.
  
\ack 
The research of IDG is co-financed by Greece and the European Union (European Social Fund- ESF) through the Operational Programme ``Human Resources Development, Education and Lifelong Learning" in the context of the project ``Strengthening Human Resources Research Potential via Doctorate Research - 2nd Cycle" (MIS-5000432), implemented by the State Scholarships Foundation (IKY). This work was supported by the Estonian Research Council grants MOBJD381, MOBTT5, MOBTT86 and PRG1055 and by the EU through the European Regional Development Fund CoE program TK133 ``The Dark Side of the Universe". TDP acknowledges the support of the grant 19-03950S of Czech Science Foundation (GA\v{C}R). The research work of VCS was supported by the Hellenic Foundation for Research and Innovation (H.F.R.I.) under the ``First Call for H.F.R.I. Research Projects to support Faculty members and Researchers and the procurement of high-cost research equipment grant'' (Project Number: 824). 

\section*{References}
\bibliography{References}{}

\end{document}